\RequirePackage{tikz}
\documentclass[sn-mathphys,iicol,sort&compress]{sn-jnl}

\usepackage{amssymb}
\usepackage{amsthm}
\usepackage{amsmath}
\usepackage{mathtools}
\usepackage{siunitx}
\usepackage{graphicx}
\graphicspath{{images}} 
\usepackage[font=small,labelfont=bf,labelsep=endash,format=plain]{caption}
\usepackage{subcaption}

\usepackage{tikz}					
\usetikzlibrary{matrix, shapes.geometric, arrows, positioning, external, decorations.fractals,%
	decorations.shapes,%
	decorations.text,%
	decorations.pathmorphing,%
	decorations.pathreplacing,%
	decorations.footprints,%
	decorations.markings}
\tikzstyle{io} = [trapezium, trapezium left angle=70, trapezium right angle=110, minimum width=3cm, minimum height=1cm, text centered, draw=black, fill=blue!30]
\tikzstyle{process} = [rectangle, minimum width=3cm, minimum height=1cm, text centered, text width=6.5cm, draw=black, rounded corners=0.2cm]
\tikzstyle{process2} = [rectangle, minimum width=3cm, minimum height=1cm, text centered, text width=3cm, draw=black, rounded corners=0.2cm]
\tikzstyle{decision} = [diamond, minimum width=3cm, minimum height=1cm, text centered, draw=black, fill=green!30]
\tikzstyle{arrow} = [thick,->,>=stealth]
\usepackage{varwidth}
\newcommand{\umbruch}[2][3cm]{\begin{varwidth}{#1}\centering#2\end{varwidth}}
\usepackage{doi}					

\usepackage{xspace}
\usepackage[]{url} 					
\usepackage[]{hyperref}
\hypersetup{
	colorlinks,
}
\usepackage[nameinlink,capitalize]{cleveref}
\usepackage{float}

\usepackage{multirow}
\usepackage{booktabs}
\usepackage{tabularx,ragged2e}
\newcolumntype{C}{>{\Centering}X}    
\usepackage{array}
\usepackage{longtable}
\usepackage{rotating}

\usepackage{csquotes}
\usepackage{xpatch}
\usepackage{lmodern}
\usepackage[obeyFinal]{todonotes}
\usepackage{chemformula}
\usepackage{enumitem}

\usepackage{pgf}
\usepackage{adjustbox}
\usepackage{setspace}

\renewcommand{\phi}{\varphi}
\renewcommand{\epsilon}{\varepsilon}

\def\etal{et~al.\xspace} 
\def\eg{e.g.,\xspace} 
\def\ie{i.e.,\xspace} 
\def\cf{cf.\xspace} 
\def\vs{vs.\xspace} 

\newcommand{\flp}{\emph{Flying Laptop}\xspace}

\renewcommand{\v}[1]{\boldsymbol{#1}}

\jyear{2023}%

\theoremstyle{thmstyleone}%
\theoremstyle{thmstyletwo}%

\theoremstyle{thmstylethree}%

\begin{document}

\title[CA Drag FLP]{Analysis of collision avoidance manoeuvres using aerodynamic drag for the Flying Laptop satellite}
\author*[1]{\fnm{Fabrizio} \sur{Turco}}\email{turcof@irs.uni-stutgart.de}

\author[1]{\fnm{Constantin} \sur{Traub}}

\author[1]{\fnm{Steffen} \sur{Gaißer}}

\author[1]{\fnm{Jonas} \sur{Burgdorf}}

\author[1]{\fnm{Sabine} \sur{Klinkner}}

\author[1]{\fnm{Stefanos} \sur{Fasoulas}}

\affil[1]{\orgdiv{Institute of Space Systems}, \orgname{University of Stuttgart}, \orgaddress{\street{Pfaffenwaldring 29}, \city{70569 Stuttgart}, \country{Germany}}}


\abstract{
	\begin{spacing}{1} 
	
	Collision avoidance is a topic of growing importance for any satellite orbiting Earth.
	Especially those satellites without thrusting capabilities face the problem of not being able to perform impulsive collision avoidance manoeuvres. For satellites in Low Earth Orbits, though, perturbing accelerations due to aerodynamic drag may be used to influence their trajectories, thus offering a possibility to avoid collisions without consuming propellant.
	Here, this manoeuvring option is investigated for the satellite \flp of the University of Stuttgart, which orbits the Earth at approximately $ \SI{600}{\kilo\meter} $.

	In a first step, the satellite is aerodynamically analysed making use of the tool ADBSat. 
	By employing an analytic equation from literature, in-track separation distances can then be derived following a variation of the ballistic coefficient through a change in attitude. A further examination of the achievable separation distances proves the feasibility of aerodynamic collision avoidance manoeuvres for the \flp for moderate and high solar and geomagnetic activity. The predicted separation distances are further compared to flight data, where the principle effect of the manoeuvre on the satellite trajectory becomes visible.
	The results suggest an applicability of collision avoidance manoeuvres for all satellites in comparable and especially in lower orbits than the \flp, which are able to vary their ballistic coefficient.
	\end{spacing}
}


\maketitle

\section{Introduction}\label{sec:introduction}
The number of objects in orbit around Earth is continuously growing. Especially in Low Earth Orbits (LEO), \ie altitudes below \SI{1500}{\kilo\meter} \cite{Montenbruck.2001}, this leads to more frequent collision warnings for functional satellites. Potential collisions do not only pose a threat to the involved functional satellites but might eventually trigger an 
avalanche-like process, rendering the LEO regime useless for future generations \cite{Kessler.1991}. 
One option for functional satellites to minimize the collision risk is the implementation of collision avoidance manoeuvres (CAMs) in case of a predicted close encounter with another object. 
Typically, such manoeuvres are performed with impulsive thrusters to deflect the satellite trajectory. Satellites without thrusting capabilities need other strategies, though, to evade potential collisions. In LEO, aerodynamic drag and lift due to the remaining atmosphere represent significant natural perturbing forces. Using them to control and manoeuvre asymmetrically-shaped satellites has been widely researched, \eg for satellite formation flight \cite{Traub.2020}. 
Although achievable accelerations are magnitudes smaller than what is possible with chemical thrusters, given enough time satellite orbits can be measurably altered, which allows for the implementation of collision avoidance manoeuvres. \par

The \flp of the University of Stuttgart is a LEO satellite launched in 2017. It frequently receives collision warnings but so far the operators were not able to take any actions to minimize the risk.\par

The following work is based on the corresponding author's master thesis. After fundamental concepts are revised in \cref{sec:fundamentals}, the \flp is aerodynamically analysed in \cref{sec:aero_analysis}. In \cref{sec:analyses}, previous collision warnings for the \flp are examined. Based on this, the feasibility of collision avoidance manoeuvres using aerodynamic drag is assessed. Lastly, analytical results for the separation distance are compared to flight data.

\subsection{Flying Laptop}
\begin{figure}[t]
	\centering
		\centering
		\includegraphics[width=0.9\linewidth]{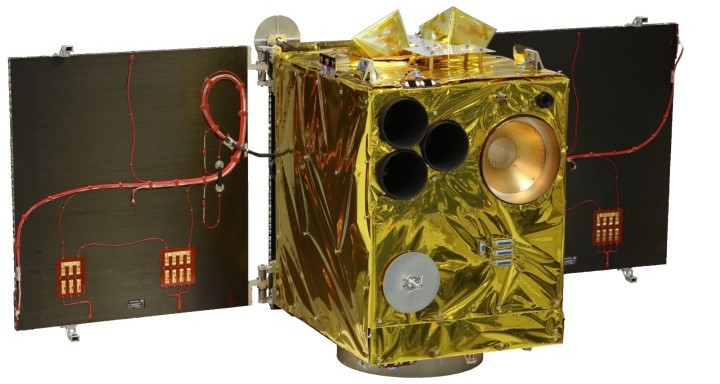}
	\caption{The \flp{} satellite of the Institute of Space Systems, University of Stuttgart.}
	\label{fig:flp}
\end{figure}

The \flp{} is the first satellite of the small satellite program of the Institute of Space Systems (IRS) at the University of Stuttgart. It was designed, manufactured and is now operated by the IRS in cooperation with partners from industry as the first prototype of the \enquote{Future Low-Cost Platform}. It takes up a volume of  \SI{0.357}{\cubic\meter} in launch configuration (\ie with contracted solar panels) and a mass of slightly more than \SI{100}{\kilogram}. It features GPS receivers, star trackers, magnetometers, gyros, and sun sensors for attitude determination and is 3-axes stabilized with magnetorquers and reaction wheels. The necessary power is generated by three solar panels, of which two are deployable \cite{Eickhoff.2016,FLP.2022}. These can be seen in \cref{fig:flp}. On July 14, 2017, it was launched into a nearly circular polar orbit with an altitude of $ \SI{\sim600}{\kilo\meter} $. \cref{tab:flp_orbit} shows the orbital parameters of the \flp{} as of April 1, 2022.\par
As a university satellite, the \flp{} is used primarily for teaching and research purposes. Further mission goals include technology demonstration and scientific Earth observation. The payloads are a multi-spectral imaging camera system (MICS), a panoramic camera (PAMCAM) and an AIS (automatic identification system) receiver for the detection of signals transmitted by ships. An experimental optical communication link can further be established for research purposes \cite{FLP.2022}.\par

\begin{table}[h]
	\centering
	\caption{The \flp{}'s mean orbital elements as of April 1, 2022 (obtained from a TLE retrieved from \url{space-track.org}). The mean anomaly is not included.}
	\label{tab:flp_orbit}
	\begin{tabular}{@{}lcc@{}} 
		\toprule
		Semi-major axis & $ a $ & \SI{6971}{\kilo\meter} \\
		Eccentricity & $ e $ & \SI{0.001228}{} \\
		Inclination & $ i $ & \SI{97.43}{\degree} \\
		Right ascension of ascending node & $ \Omega $ & \SI{310.2}{\degree} \\
		Argument of perigee & $ \omega $ & \SI{266.0}{\degree} \\
		\bottomrule
	\end{tabular}
\end{table}

\section{Fundamentals}\label{sec:fundamentals}

\subsection{Satellite aerodynamics}
\label{sec:satellite_aerodynamics}
\begin{figure}[h]
	\centering
	\begin{tikzpicture}
		\node[rectangle, draw, fill=gray, opacity=1, minimum width=1cm, minimum height=0.5cm, rotate around={45:(0,0)}]{} (0,0);
		\draw[black,fill=black] (0,0) circle (.5ex);
		\draw[arrows={-latex}]  (0,0) -- (3,0) node[midway,anchor=north west]{$\boldsymbol{v}_{rel}$};
		\draw[arrows={-latex}]  (0,0) -- (-2,0) node[midway,anchor=north]{$\boldsymbol{f}_D$};
		\draw[arrows={-latex}]  (-2,0) -- (-2,0.5) node[midway,anchor=east]{$\boldsymbol{f}_L$};
		\draw[arrows={-latex}]  (0,0) -- (-2,0.5) node[midway,anchor=south]{$\boldsymbol{f}_A$};
	\end{tikzpicture}
	\caption{The specific aerodynamic force $ \v{f}_A $ of a satellite is separated into a lift component $ \v{f}_L $ and a drag component $ \v{f}_D $.}
	\label{fig:aero_forces}
\end{figure}
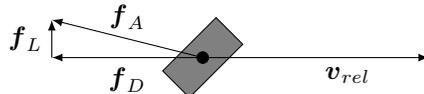

The remaining atmospheric particles in LEO result in a force acting on any satellite. Similarly to aircraft aerodynamics, the aerodynamic force can be divided into a lift and a drag component, as shown in \cref{fig:aero_forces}.

The specific lift force $ \v{f}_L $ lies within the plane perpendicular to the relative velocity and will not be further considered, as lift coefficients are usually small for satellites.
The specific drag force $ \v{f}_D $ acts in anti-parallel direction to the relative velocity of the local atmosphere and is defined as \cite{Vallado}
\begin{equation}
	\v{f}_D = -\frac{1}{2} \rho \frac{C_D A_{ref}}{m} v_{rel}^2 \frac{\v{v}_{rel}}{v_{rel}}.
	\label{eq:drag}
\end{equation}
Here, $ C_D $ is the satellite's drag coefficient, $ A_{ref} $ is the reference area, $ m $ denotes the satellite mass, $ \rho $ is the atmospheric density and the satellite's velocity relative to the local atmosphere is denoted as $ \v{v}_{rel} $. Neglecting atmospheric winds, the relative velocity equals the satellite's orbital velocity. The reference area can be arbitrarily chosen. For this work, the projected satellite area in the direction of flight, \ie the effective cross-section, is used.

The ballistic coefficient $ \beta $ and its inverse $\beta^*$ (IBC) are defined as \cite{Vallado}
\begin{equation}
	\beta = \frac{m}{C_D A_{ref}},
\end{equation}
\begin{equation}
	\beta^* = \frac{C_D A_{ref}}{m}.
	\label{eq:ballistic_coeff}
\end{equation}
Satellites with higher IBC experience higher specific drag forces than such with low IBC.\par

Regarding the specific drag force, the only option for satellite control lies within a variation of the IBC, either by varying the effective satellite cross-section or by changing the drag coefficient. All other parameters are either fixed or out of operator control.

%

\subsubsection{Sentman model}
\label{sec:sentman_model}
Conditions in LEO lead to a flow regime called free-molecular flow. Aerodynamic interaction is here driven only by collisions between the atmospheric particles and the spacecraft, as collisions between particles are very rare. Complex phenomena take place during the interaction of the particles with the satellite's surface. To describe them, several gas-surface interaction models have been developed \cite{Livadiotti.2020c}. Most widely used for drag estimations at comparatively low altitudes is the Sentman model \cite{ADBSatMethodology.2022,Moe.2005, Sentman.1961}.\par
\begin{figure}[b]
	\centering
	\includegraphics[width=\linewidth]{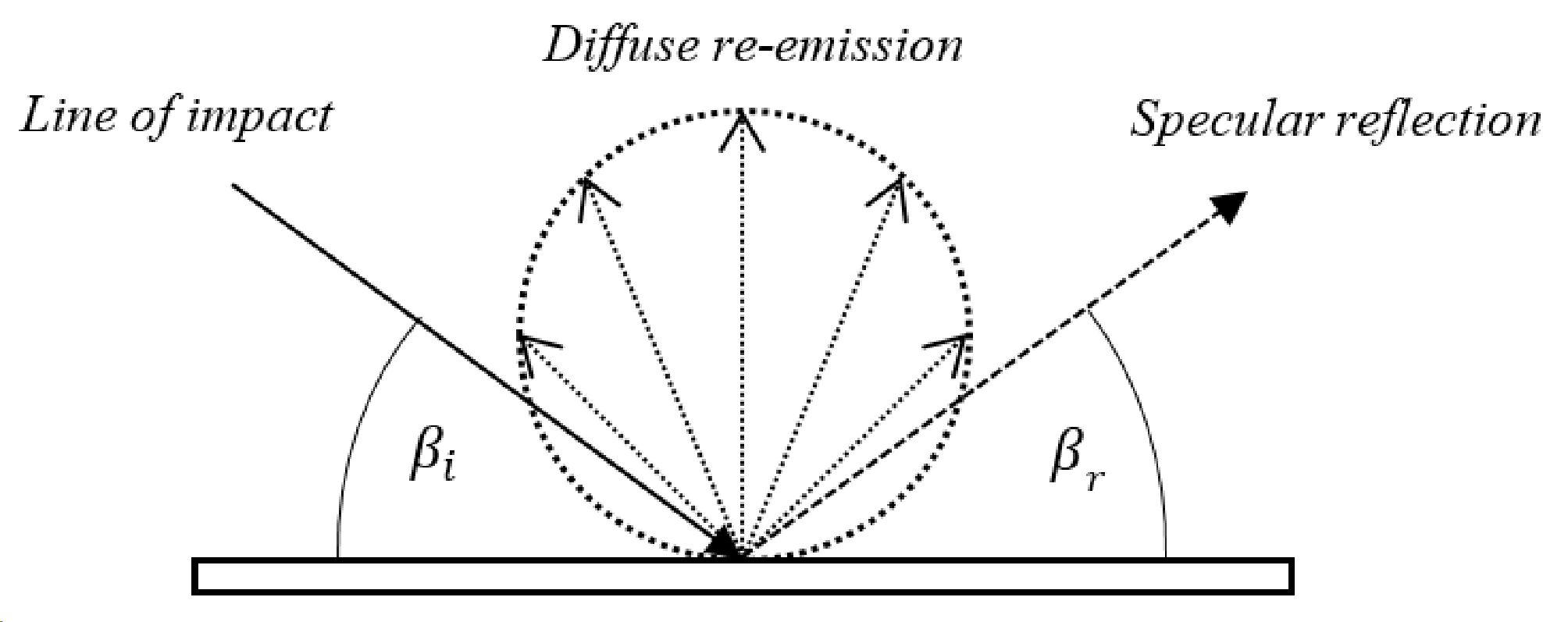}
	\caption{Specular reflection \vs diffuse re-emission \cite{Traub.2020}. The angle of incidence is $ \beta_i $, the angle of reflection is $ \beta_r $.}
	\label{fig:reflectiontypes}
\end{figure}

When a particle hits the satellite's surface, its reflection can happen in two extreme ways depending on the interactions: Either, it can be rejected under the same angle as it hit the surface, which is called specular reflection - very much like light is reflected by a mirror. Or it can be re-emitted independently from its incident angle during the other extreme case of diffuse reflection. The direction and velocity of the rejection is distributed in a probabilistic way depending on the wall temperature (\cf \cref{fig:reflectiontypes}).

Further, the Sentman model makes use of an accommodation coefficient $ \alpha_T $ to describe energy transfer from particles to the wall:
\begin{equation}
	\alpha_T = \frac{E_i-E_r}{E_i-E_w} = \frac{T_{k,i}-T_{k,r}}{T_{k,i}-T_w}.
	\label{eq:accom_coeff}
\end{equation}
It is defined as the quotient of the differences in energy between the incoming particles $ E_i $ and reflected particles $ E_r $ and between the incoming particles and the wall $ E_w $. This can be equally described by a quotient of differences in temperature $ T $. Sentman further defined momentum accommodation coefficients. They specify the ratio of the difference in momentum between the incoming (\enquote{$ i $}) and reflected particles (\enquote{$ r $}) and between the incoming particles and the wall (\enquote{$ w $}). They are defined for tangential momentum $ \tau $ and normal momentum $ p $:
\begin{align}
	\sigma_t &= \frac{\tau_i - \tau_r}{\tau_i - \tau_w} = \frac{\tau_i - \tau_r}{\tau_i},\\
	\sigma_n &= \frac{p_i - p_r}{p_i - p_w}.
\end{align}
Per definition of diffuse reflection $ \tau_w=0 $ \cite{Sentman.1961}.\par
Following the previous explanations, for specular reflection without thermal accommodation $ \sigma_t=\sigma_n=0 $. In case of completely diffuse reflection and full thermal accommodation, the particles reach thermal equilibrium with the surface and $ \sigma_t=\sigma_n=1 $. The Sentman model assumes diffuse reflection and a variable thermal accommodation coefficient \cite{Traub.2020}. Therefore, $ \sigma_t=1 $ and $ \sigma_n $ depends on the degree of thermal accommodation.\par
Taking into account the thermal motion of the particles as well as the angle of incidence, the wall temperature and the temperature of the incoming particles, the Sentman model allows the calculation of a drag and a lift coefficient for a surface in free molecular flow. 

The model has been implemented in the {ADBSat} tool developed by \emph{Sinpetru}, \emph{Crisp} \etal, which calculates a satellite's drag coefficient using a CAD model as input. The satellite geometry is divided into different panels, \ie flat plates, and the drag coefficient is evaluated for each of them. The individual panels' contributions are added to form the total drag coefficient, considering potential shading of some panels \cite{ADBSatMethodology.2022}.

\subsection{Conjunctions}
\label{sec:conjunctions}
A conjunction is a close encounter between a satellite and another object (a defunct or working satellite or debris object). Such conjunctions pose an inherent threat because of the risk of a potential collision, which almost certainly results in the loss of the satellite. Moreover, collisions produce a high amount of debris particles, which themselves increase the collision risk for other operating satellites. This section describes how conjunction warnings are issued and how collision probabilities can be calculated.\par

The Joint Space Operations Centre (JSpOC) provides services for satellite operators regarding conjunction events \cite{NASA.2020,SafetyHandbook.2020}. 
A proprietary catalogue of satellites and debris objects is continuously screened for close encounters with a miss distance below a defined threshold. When a close encounter is identified, more thorough analyses are performed including high-fidelity orbit determination and propagation as well as uncertainty estimations. The available information regarding a conjunction is then summarized in a Conjunction Data Message (CDM), which is sent to the satellite operators. The CDM's format and content are standardized \cite{CCSDSCDM.2013}.
Each CDM contains information about the two objects involved in the encounter, including (if applicable) name, type, operator name, whether they are manoeuvrable or not and their considered radius. The predicted time of closest approach (TCA) and the predicted positions and velocities of both objects are presented, accompanied by an estimate of the position covariance matrices. The observation time intervals and the used models for orbit determination and subsequent propagation are stated as well as the IBC and solar radiation coefficient used for propagation. The stated IBC is to be highlighted because it will be used as a reference value later on. Lastly, orbital information about the objects is given, including apogee and perigee heights, eccentricity, and inclination.

\subsection{Analysis tool for collision avoidance manoeuvres using aerodynamic drag}
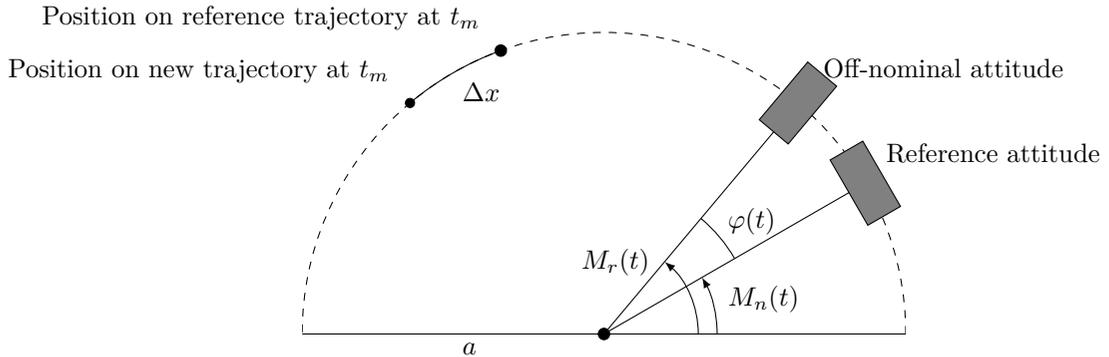
\begin{figure*}[hbt!]
	\centering
	\begin{tikzpicture}
		\draw[black,fill=black] (0,0) circle (.5ex);
		\draw (-4,0) -- (0,0) node[midway,anchor=north west]{$a$};
		\draw (0,0) -- (4,0);
		\begin{scope}
			\clip (-4,0) rectangle (4,4);
			\draw[dashed] (0,0) circle(4);
		\end{scope}
		\draw (0,0) -- (50:4)
		node[rectangle, draw, fill=gray, opacity=1, minimum width=1cm, minimum height=0.5cm, rotate=50] (a) {}
		node[right=1.4ex,yshift=1.4ex, anchor=south west] {Off-nominal attitude};
		\draw[arrows={-latex}] (1.25,0) arc (0:50:1.25) node[left=0.5ex]{$M_r(t)$};
		
		\draw (0,0) -- (30:4) 
		node[rectangle, draw, fill=gray, opacity=1, minimum width=1cm, minimum height=0.5cm, rotate=120] (b) {}
		node[right=1ex,yshift=1ex, anchor=south west] {Reference attitude};
		\draw[arrows={-latex}] (1.5,0) arc (0:30:1.5) node[midway,right=0.5ex,yshift=0.5ex]{$M_n(t)$};
		
		\draw[] (30:2) arc (30:50:2) node[midway,right=-0.1ex,yshift=1.4ex]{$\phi (t)$};
		
		\draw[black,fill=black, radius=.4ex] (110:4) circle(.5ex) node[left=1ex,yshift=1ex,anchor=south east] {Position on reference trajectory at $ t_m $};
		\draw[black,fill=black, radius=.4ex] (130:4) circle(.4ex) node[left=1ex,yshift=1ex,anchor=south east] {Position on new trajectory at $ t_m $};
		
		\draw[] (110:4) arc (110:130:4) node[midway,anchor=north west]{$\Delta x$};
	\end{tikzpicture}
	\caption{Concept of a collision avoidance manoeuvre using aerodynamic drag \cite{Turco.2023}. The off-nominal attitude has an increased inverse ballistic coefficient, leading to a positive in-track separation distance to the reference trajectory. The satellites started from equal initial conditions and the separation distance builds up over time.}
	\label{fig:manoeuvre_schematic}
\end{figure*}

Changing a satellite's ballistic coefficient for a specified duration alters the experienced drag force and therefore the satellite's trajectory. As drag acts in the orbital plane, the effect of the orbit can be approximated as an in-track separation distance relative to the trajectory if the ballistic coefficient were not varied. \cref{fig:manoeuvre_schematic} schematically presents the influence of an increase in IBC. The resulting increased drag force leads to the satellite advancing relative to its original trajectory. This is due to a negligibly small decrease of the satellite's semi-major axis, which results in a decrease in the orbital period.\par
Omar and Bevilacqua \cite{Omar.2020} derived an analytic equation for estimating achievable separation distances and proposed its use for collision avoidance. The separation distance $ \Delta x $ depends on the average atmospheric density encountered during the manoeuvre $ \bar \rho $, on the semi-major axis $ a $, on the change in IBC $ \Delta \beta^* $ and on the manoeuvre duration $ t_m $:
\begin{align}
	\Delta x = \frac{3\bar \rho \mu_E}{4a} \Delta \beta^* t_m^2
	\label{eq:deltax_min}\\
	\text{with\quad} \Delta \beta^* = \left( \beta^* - \beta^*_{ref} \right).
\end{align}
$ \Delta \beta^* $ is the change in IBC in the commanded attitude $ \beta^* $ compared to the reference value $ \beta^*_{ref} $. $ \mu_E $ is the gravitational parameter of the Earth.

Based on this, an analysis tool for collision avoidance manoeuvres using aerodynamic drag has been developed \cite{Turco.2022,Turco.2023}. It is based on the analytical equation and estimates achievable in-track separation distances through a change in IBC. It considers recent space weather data for atmospheric models as well as density estimation and assesses the influence of potential manoeuvres on the collision risk. 

\section{Aerodynamic analysis of the \flp}\label{sec:aero_analysis}
To estimate the influence of changes in attitude on the \flp's trajectory, the satellite is analysed with regard to its aerodynamic properties. For this, an available software tool is used.

\subsection{Model of the \flp{}}
\label{sec:aero_model}
\begin{figure}[h!]
	\centering
	\begin{subfigure}{0.475\linewidth}
		\centering
		\includegraphics[width=\textwidth]{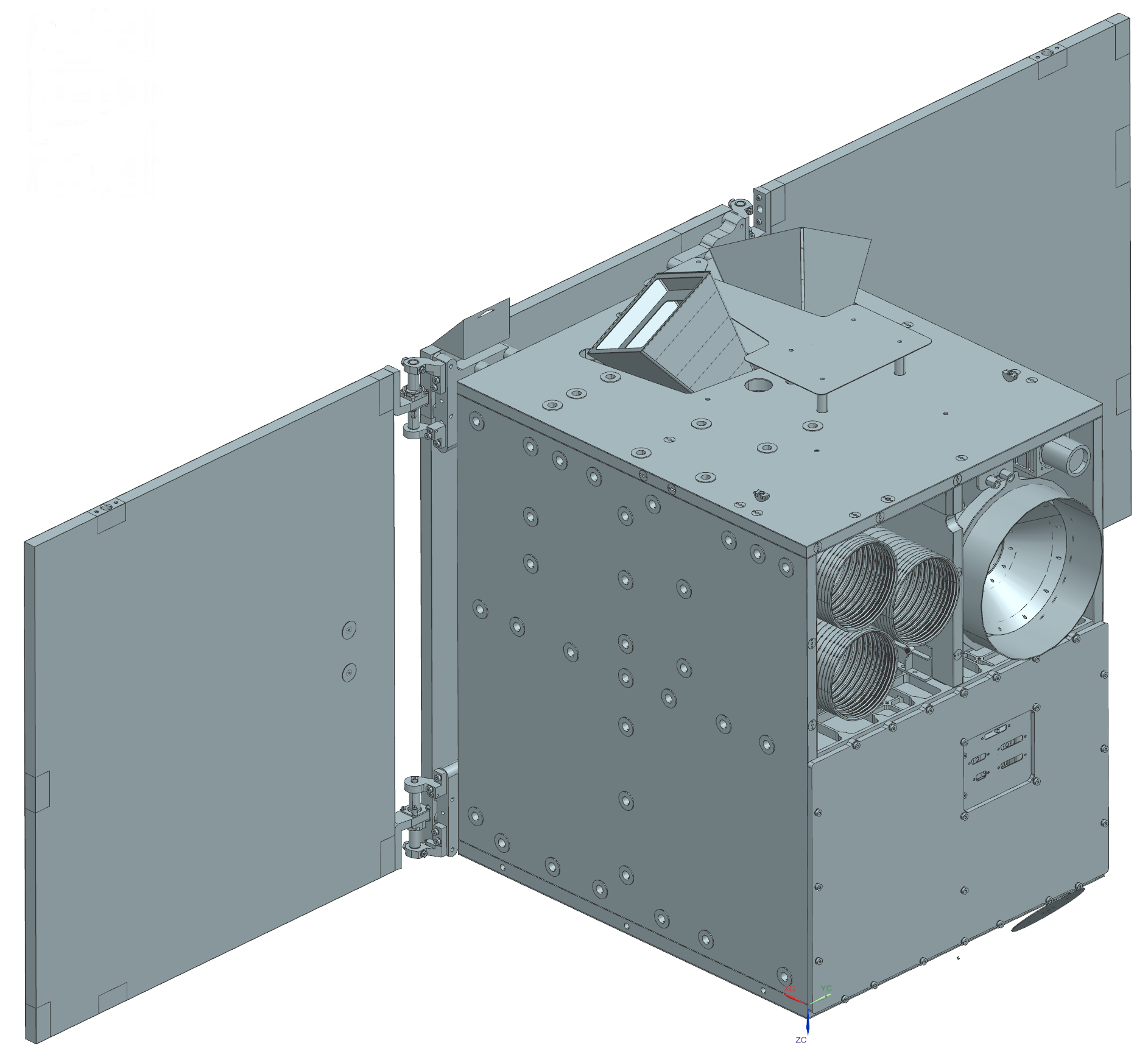}
		\caption{Detailed model.}
	\end{subfigure}
	\hfill
	\begin{subfigure}{0.475\linewidth}
		\centering
		\includegraphics[width=\textwidth]{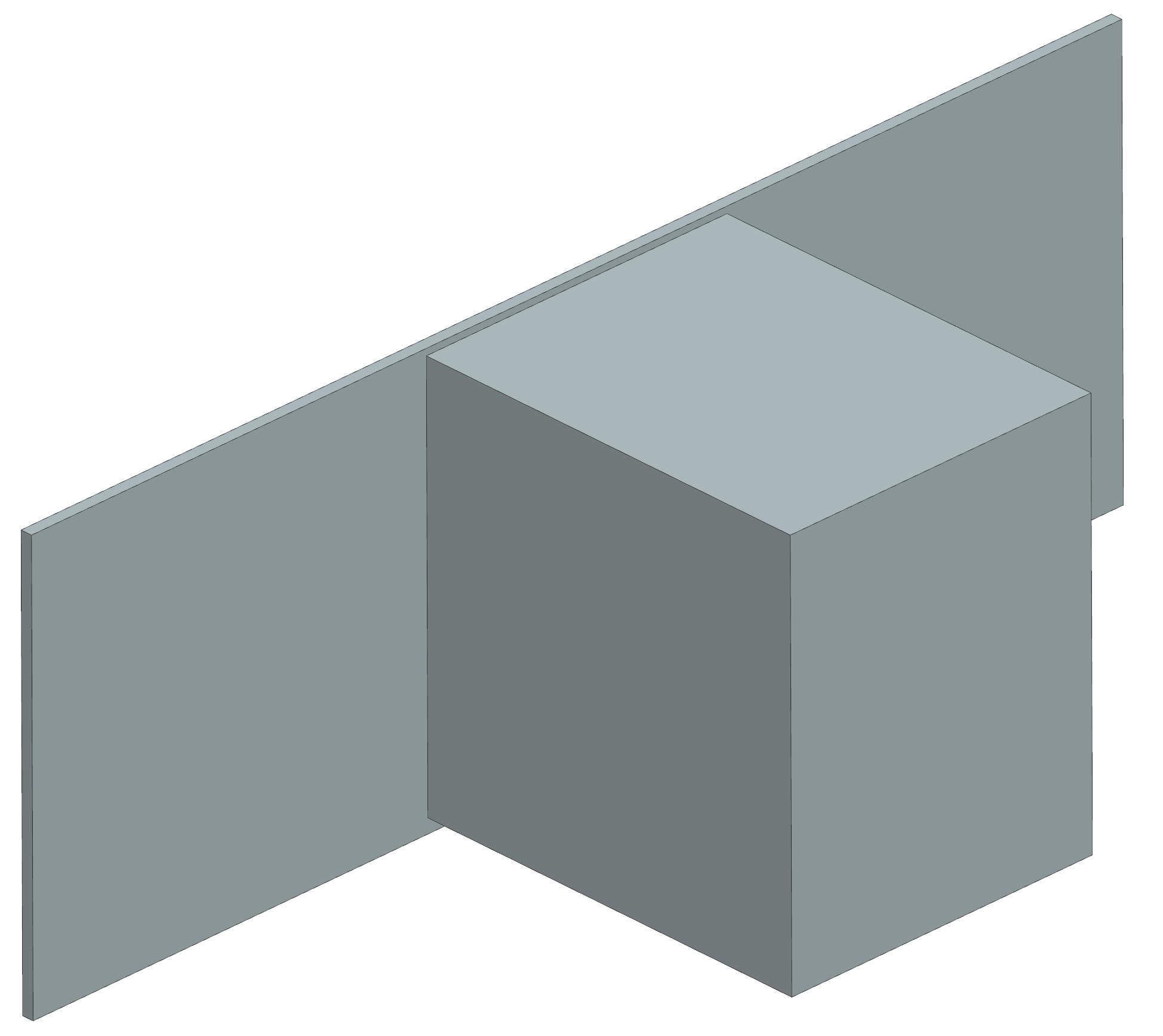}
		\caption{Simplified model.}
	\end{subfigure}
	\caption{Comparison of the detailed CAD model and the simplified model used for the aerodynamic analysis.}
	\label{fig:aero_models}
\end{figure}

To decrease computational costs, the original CAD model of the \flp is simplified, leading to a drastic decrease in individual surfaces. A comparison of the original and the simplified model is shown in \cref{fig:aero_models}. The major surfaces are still accounted for in the simplified model, while the instruments are removed. The main body of the satellite bus measures approximately $ 530 \times 670 \times \SI{660}{\milli\meter\cubed}$ ($x\times y\times z$), while the solar panels have a span of about $ \SI{1922}{\milli\meter} $. The weight is $ \SI{108.8}{\kilogram} $.\par

\begin{figure}[h]
	\centering
	\begin{subfigure}[c]{0.3\linewidth}
		\centering
		\resizebox{\linewidth}{!}{
			\begin{tikzpicture}[baseline=(current bounding box.center)]
				\node[rectangle, draw, fill=gray, opacity=1, minimum width=1.922cm*1.5, minimum height=0.67cm*1.5, rotate around={90:(0,0)}]{} (0,0);
				\node[rectangle, draw, fill=gray, opacity=1, minimum width=0.53cm*1.5, minimum height=0.67cm*1.5, rotate around={90:(0,0)}]{} (0,0);
				\draw[arrows={-latex}]  (0,0) -- (2,0) node[anchor=north west]{$\v{v}$};
				\draw[arrows={-latex},blue]  (0,0) -- (0,1) node[anchor=north east]{$\v{x}$};
				\draw[arrows={-latex},blue]  (0,0) -- (-1,0) node[anchor=north east]{$\v{y}$};
		\end{tikzpicture}}
		\caption{Minimum drag attitude.}
		\label{fig:minimum_drag_attitude}
	\end{subfigure}\hfill
	\begin{subfigure}[c]{0.3\linewidth}
		\centering
		\resizebox{\linewidth}{!}{
			\begin{tikzpicture}[baseline=(current bounding box.center)]
				\node[rectangle, draw, fill=gray, opacity=1, minimum width=1.922cm*1.5, minimum height=0.67cm*1.5, rotate around={0:(0,0)}]{} (0,0);
				\node[rectangle, draw, fill=gray, opacity=1, minimum width=0.53cm*1.5, minimum height=0.67cm*1.5, rotate around={0:(0,0)}]{} (0,0);
				\draw[arrows={-latex}]  (0,0) -- (2,0) node[anchor=north west]{$\v{v}$};
				\draw[arrows={-latex},blue]  (0,0) -- (1,0) node[anchor=north east]{$\v{x}$};
				\draw[arrows={-latex},blue]  (0,0) -- (0,1) node[anchor=north east]{$\v{y}$};
				
				\vphantom{
					\node[rectangle, draw, fill=gray, opacity=0, minimum width=1.922cm*1.5, minimum height=0.67cm*1.5, rotate around={90:(0,0)}]{} (0,0);
					\node[rectangle, draw, fill=gray, opacity=0, minimum width=0.53cm*1.5, minimum height=0.67cm*1.5, rotate around={90:(0,0)}]{} (0,0);
					\draw[arrows={-latex}]  (0,0) -- (2,0) node[anchor=north west]{$\v{v}$};
					\draw[arrows={-latex},blue]  (0,0) -- (0,1) node[anchor=north east]{$\v{x}$};
					\draw[arrows={-latex},blue]  (0,0) -- (-1,0) node[anchor=north east]{$\v{y}$};
				}
		\end{tikzpicture}}
		\caption{Nadir-pointing attitude.}
		\label{fig:Nadir_pointing_attitude}
	\end{subfigure}\hfill
	\begin{subfigure}[c]{0.3\linewidth}
		\centering
		\resizebox{\linewidth}{!}{
			\begin{tikzpicture}[baseline=(current bounding box.center)]
				\draw[fill=gray, opacity=1, minimum width=0.53cm*1.5, minimum height=0.66cm*1.5,]{} (-0.33*1.5,-0.265*1.5) rectangle ++(0.66*1.5,0.53*1.5);
				\draw[fill=gray, opacity=1, minimum width=0.53cm*1.5, minimum height=0.66cm*1.5,]{} (0.33*1.5-0.015*1.5,-0.961*1.5) rectangle ++(0.03*1.5,1.922*1.5);
				\draw[arrows={-latex}]  (0,0) -- (2,0) node[anchor=north west]{$\v{v}$};
				\draw[arrows={-latex},blue]  (0,0) -- (0,-1) node[anchor=north east]{$\v{x}$};
				\draw[arrows={-latex},blue]  (0,0) -- (-1,0) node[anchor=north east]{$\v{z}$};
		\end{tikzpicture}}
		\caption{Maximum drag attitude.}
		\label{fig:maximum_drag_attitude}
	\end{subfigure}
	\caption{Schematic illustrations of the different specified attitudes of the \flp with the body coordinate system. All attitudes can be rotated around the direction of flight without changing the flow field.}
	\label{fig:attitudes}
\end{figure}
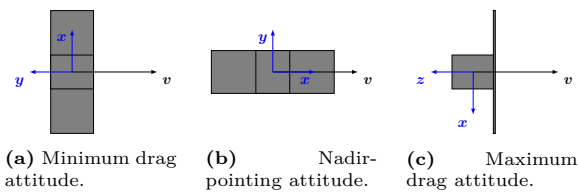

The IBC of the \flp can be maximized by maximizing the area perpendicular to the relative velocity. When no atmospheric wind is assumed, the relative velocity is parallel to the direction of flight. This corresponds to a flight with maximum drag. Vice versa, the minimization of this area leads to a minimum IBC and minimum drag. The respective maximum and minimum drag attitudes are defined by having either the body $ z $-axis (maximum drag) or $ y $-axis (minimum drag) point parallel to the direction of flight. To protect the payload cameras and the star trackers from atmospheric particles, the negative $ z $ or $ y $-axes are pointed towards flight direction. The Nadir-pointing attitude is defined by having the $ z $-axis pointing towards the centre of Earth and the $ x $-axis in the direction of flight. The different attitudes are shown in \cref{fig:attitudes}.

\subsection{ADBSat analysis}
ADBSat offers the possibility to analyse a satellite geometry with regards to its aerodynamic properties \cite{ADBSatMethodology.2022}. Besides the CAD model of the satellite, a position (latitude, longitude and altitude), time and the respective indices for solar and geomagnetic activity for the NRLMSISE-00 atmosphere model are needed as inputs. The NRLMSISE-00 model is used to derive the atmospheric composition and temperature at the position and time. Based on this, the gas-surface interactions are modelled using the Sentman model. Finally, the aerodynamic force on every surface panel is calculated. The individual panels' contributions are added to form the total aerodynamic forces and coefficients. Since the IBC is, therefore, dependent on atmospheric composition and temperature, it will be analysed for varying solar and geomagnetic activity levels.\par
As outlined in \cref{sec:sentman_model}, different parameters are needed as input for the Sentman model. 
The wall temperature is set to be $ \SI{300}{\kelvin} $, as it is often assumed in literature \cite{Doornbos.2011,Traub.2020_2}. The accommodation coefficient requires closer examination.

\subsubsection{Accommodation coefficient}
Satellite aerodynamics and gas-surface interaction have been studied intensively for Very Low Earth Orbits (VLEO), \ie orbits below $ \SI{450}{\kilo\meter} $ \cite{Crisp.2020}, including the development of several models. Still, less is known about altitude regimes above $ \SI{500}{\kilo\meter} $ and the applicability of the aforementioned models has to be checked carefully. Regarding the introduced Sentman model, this especially concerns the accommodation coefficient.
In literature, the Semi-Empirical Satellite Accommodation Model (SESAM) is often used for estimating the accommodation coefficient depending on the number density of atomic oxygen $ n_{\ch{O}} $ and the atmospheric temperature $ T $ \cite{Traub.2020_2}:
\begin{equation}
	\alpha_T = \frac{7.5 \cdot 10^{-17} \cdot n_{\ch{O}} T}{1+7.5\cdot 10^{-17} \cdot n_{\ch{O}} T}.
\end{equation}
It is estimated to be valid for $ 0.85 \le \alpha \le 1 $ and to be used with the NRLMSISE-00 model to obtain the number density of atomic oxygen and the atmospheric temperature \cite{Traub.2020_2, Pilinski.2011, Moe.2005}. Alternatively, a valid altitude range of $ \SIrange{100}{500}{\kilo\meter} $ is given. This is approximately the region, in which the model's output accommodation coefficient typically satisfies $ 0.85 \le \alpha \le 1 $.
For the altitude regime of the \flp, the accommodation coefficients as calculated by SESAM can only be considered valid for high solar and geomagnetic activity.
In general, accommodation tends to decrease with increasing altitudes due to less adsorption of molecules at the surface \cite{Moe.2005}. This shows in a shift to specular reflection leading to higher aerodynamic coefficients.\par
In the absence of better estimates, the accommodation coefficient is here obtained using the SESAM model. The output is, however, restricted to be minimum $ 0.85 $ and thus valid in the sense of the model. This leads to the determined IBC representing a lower estimate since aerodynamic drag increases with decreasing accommodation levels.
For the maximum drag configuration, this corresponds to a conservative approach. For the minimum drag configuration, however, it needs to be considered that the achievable difference in IBC might be lower than calculated here.

\subsubsection{Results}
\begin{figure}[h!]
	\centering
	\begin{subfigure}{0.48\linewidth}
		\centering
		\includegraphics[trim=122 282 205 293,clip,width=\textwidth]{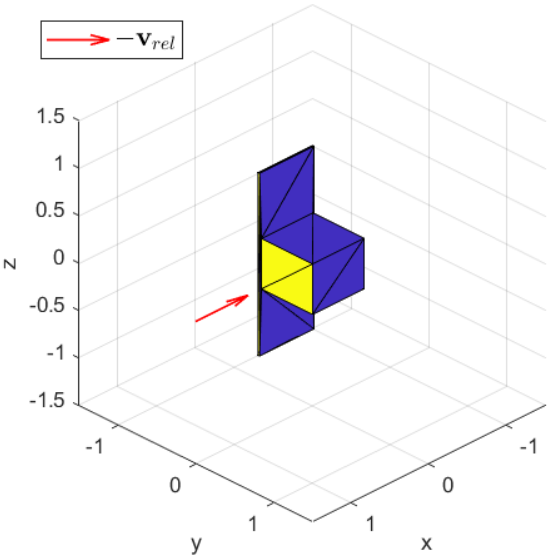}
		\caption{Minimum drag attitude.}
	\end{subfigure}\hfill
	\begin{subfigure}{0.48\linewidth}
		\centering
		\includegraphics[trim=122 282 205 293,clip,width=\textwidth]{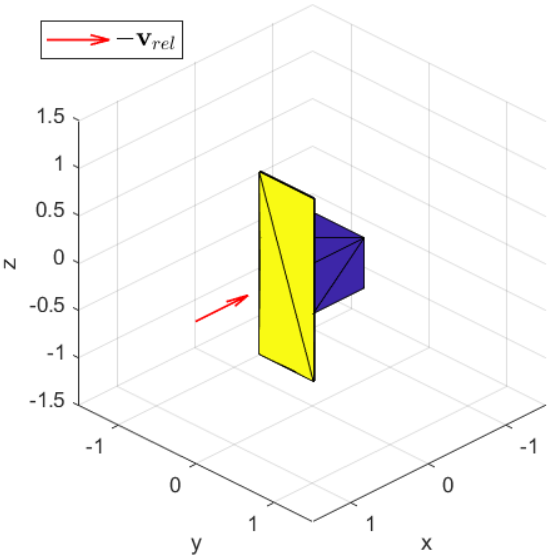}
		\caption{Maximum drag attitude.}
	\end{subfigure}
	\caption{Visualizations from the ADBSat tool for minimum and maximum drag attitude for moderate solar activity at $ \SI{600}{\kilo\meter} $ altitude over $ \SI{0}{\degree} \text{N}, \SI{0}{\degree} \text{E} $ on April 1, 2022, 12:00 UTC. Yellow areas correspond to areas highly contributing to aerodynamic drag.}
	\label{fig:aero_adbsat}
\end{figure}

The three different attitudes (\cf \cref{fig:attitudes}) are analysed with the ADBSat tool. Atmospheric conditions are heavily dependent on the geocentric position. Therefore, an approximated circular orbit of the \flp is assumed with an inclination of $ \SI{97.4}{\degree} $ and an LTAN of 06:00 \unit{\hour}. The ADBSat tool is employed to determine the drag coefficient of the satellite in the respective attitudes at 100 points on this orbit and the results are then averaged over these evaluation points. 
Using \cref{eq:ballistic_coeff}, the drag coefficients can be translated into corresponding ballistic coefficients. ADBSat uses the largest surface, \ie the solar panels, as a reference area in each attitude. \cref{fig:aero_adbsat} shows the analysis visualizations in ADBSat for minimum and maximum drag attitude. The areas exposed to the flow are the ones which mainly contribute to aerodynamic drag.

%
\cref{tab:aero_results} provides the resulting IBCs for the diffferent activity levels (\cf \cref{tab:activity_levels}). It becomes clear that the IBC is significantly lower in minimum drag attitude, indicating a lower drag force acting on the satellite. In minimum drag attitude and Nadir-pointing, the IBCs show a minimum for moderate activity and slightly higher but comparable values for high activity. The IBCs for moderate activity amount to $ \SI{87.77}{\percent} $ and $ \SI{89.81}{\percent} $ of the value for low activity, respectively. In maximum drag attitude, the IBC is less dependent on the activity level, showing a difference of only $ \SI{3.393}{\percent} $ between low and high activity.

\begin{table}[htb!]
	\centering
	\caption{Inverse ballistic coefficient $ \beta^* $ [\unit{\square\meter\per\kilogram}] of the \flp for different attitudes and different levels of solar and geomagnetic activity.}
	\label{tab:aero_results}
	\begin{tabular}{@{}lccc@{}}
		\toprule
		&\multicolumn{3}{c}{$ \beta^* $ [\unit{\square\meter\per\kilogram}]}\\
		\midrule
		Activity level&Low&Moderate&High\\
		\midrule
		Minimum drag &$ \qty{1.384e-2}{} $ &$ \qty{1.214e-2}{} $ &$ \qty{1.220e-2}{} $ \\
		Nadir-pointing &$ \qty{1.475e-2}{} $ &$ \qty{1.324e-2}{} $ &$ \qty{1.328e-2}{} $ \\
		Maximum drag &$ \qty{3.377e-2}{} $ &$ \qty{3.262e-2}{} $ &$ \qty{3.258e-2}{} $ \\
		\bottomrule
	\end{tabular}
\end{table}

%

\section{Analyses}\label{sec:analyses}

\subsection{Previous collision warnings for the \flp}\label{sec:cdm_analysis}
In this section, the conjunction warnings from JSpOC from the \flp's launch in July 2017 until September 2022 will be reviewed. In total, the \flp operators received 5114 collision warnings. 4136 of them showed a collision probability of $ P_c=0 $. This is due to many close encounters fulfilling the report conditions but leading to negligible collision probabilities. These encounters will not be included in the following investigations.
The 978 warnings with $ P_c>0 $ were issued for 177 close encounters in total, leading to an average of 5.53 CDMs per close encounter. The CDMs are typically updated regularly until the time of closest approach. Of the 177 close encounters, 167 showed a collision probability $ \ge\SI{1e-4}{} $ in at least one of the issued CDMs, which is the typical threshold for the application of a CAM. On average, such a close encounter occurs every 11.32 days for the \flp.

\begin{figure}[t]
	\centering
	\input{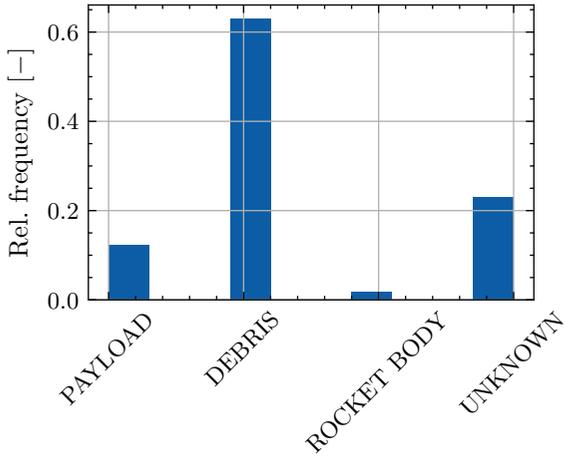}
	\caption{Types of the secondary objects in close encounters with the \flp.}
	\label{fig:type}
\end{figure}

As can be seen in \cref{fig:type}, the secondary object was identified as debris in more than $ \SI{60}{\percent} $ of the close encounters. In approximately a quarter of the cases, the \flp encountered objects of unknown type and only $ \SI{12.43}{\percent} $ close encounters took place between the \flp and another payload. Of these payloads, $ \SI{63.64}{\percent} $ were reported to be manoeuvrable. In total, only $ \SI{7.910}{\percent} $ of the secondary objects were proven to be manoeuvrable - conjunctions with objects which are themselves not capable of performing avoidance manoeuvres are the regular case. This highlights the need for the \flp's operators to take measures to decrease collision risk.

\begin{figure}[t]
	\centering
	\resizebox{\linewidth}{!}{\input{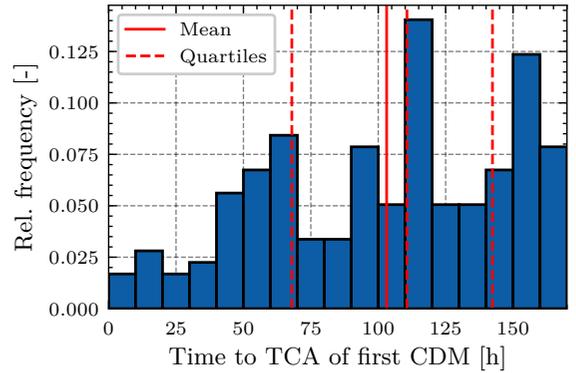}}
	\caption{Histogram of the time between creation of the first CDM and TCA for close encounters between the \flp and a secondary object.}
	\label{fig:analysis_time_allCDMs}
\end{figure}

For every individual close encounter, the time from the creation of the first CDM until the predicted TCA was determined. The CDMs are received shortly after creation. Therefore, the mentioned time interval is a measure of the available time to plan and implement CAMs. The resulting distribution is plotted in \cref{fig:analysis_time_allCDMs}. Only in $ \SI{13.00}{\percent} $ of the examined cases, the available time is $ \le \SI{48}{\hour} $, in $ \SI{29.40}{\percent} $ it is $ \le \SI{72}{\hour} $. On average, there is an interval of $ \SI{103.64}{\hour} $ between the first CDM and TCA. For $ \SI{50}{\percent} $ of the encounters, the available time is $ \ge \SI{110.69}{\hour} $. Most encounters are first discovered $ \SIrange{110}{120}{\hour} $ before the time of closest approach.

\begin{figure}[t]
	\centering
	\resizebox{\linewidth}{!}{\input{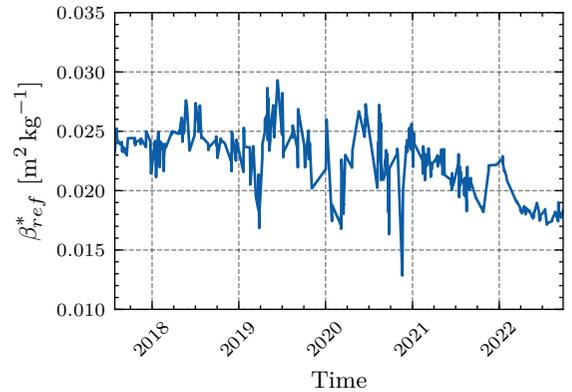}}
	\caption{The reference inverse ballistic coefficient as stated in JSpOC CDMs over time.}
	\label{fig:analysis_CB_allCDMs}
\end{figure}

Lastly, the reference IBC published in the CDMs is examined, \cref{fig:analysis_CB_allCDMs} shows its evolution. Its maximum is reached in July 2019 with a value of $ \beta^*_{ref} = \SI{0.02930}{\square\meter\per\kilogram} $, the minimum is $ \SI{0.01284}{\square\meter\per\kilogram} $ in November, 2020. The reference IBC stays within the values determined during the \flp's aerodynamic analysis (cf. \cref{tab:aero_results}).
The overall average is $ \SI{0.02238}{\square\meter\per\kilogram} $. Beginning with the start of 2021, the reference IBC starts to decrease steadily until today. Considering only CDMs since April 2022 the reference IBC shows an average of $ \SI{0.01794}{\square\meter\per\kilogram} $.

\subsection{Feasibility analysis of collision avoidance using aerodynamic drag}
With the analysis tool for CAMs using aerodynamic drag, the separation distance achievable by the \flp was analysed. The results were first published in \cite{Turco.2023}.

A reference trajectory is based on a CDM received for a close encounter on April 7, 2022. The reference IBC is $ \beta^*_{ref} = \qty{0.01794}{\square\meter\per\kilogram}$, which is the average of the previously analysed CDMs. To maximize the achievable manoeuvre effect, the minimum and maximum drag attitude of the \flp are examined, representing minimum and maximum IBC, respectively.
The separation distance relative to the reference trajectory is estimated for both attitudes and differing solar and geomagnetic activity levels, as defined in \cref{tab:activity_levels}. 
The density is evaluated and averaged over one orbital period of the reference orbit using the NRLMSISE-00 model. The resulting values are 
$ \bar \rho = \qty{1.158e-14}{\kilogram\per\cubic\meter} $ for low, 
$ \bar \rho = \qty{1.650e-13}{\kilogram\per\cubic\meter} $ for moderate and 
$ \bar \rho = \qty{1.020e-12}{\kilogram\per\cubic\meter} $ for high activity.

\begin{figure*}[p!]
	\centering
	\input{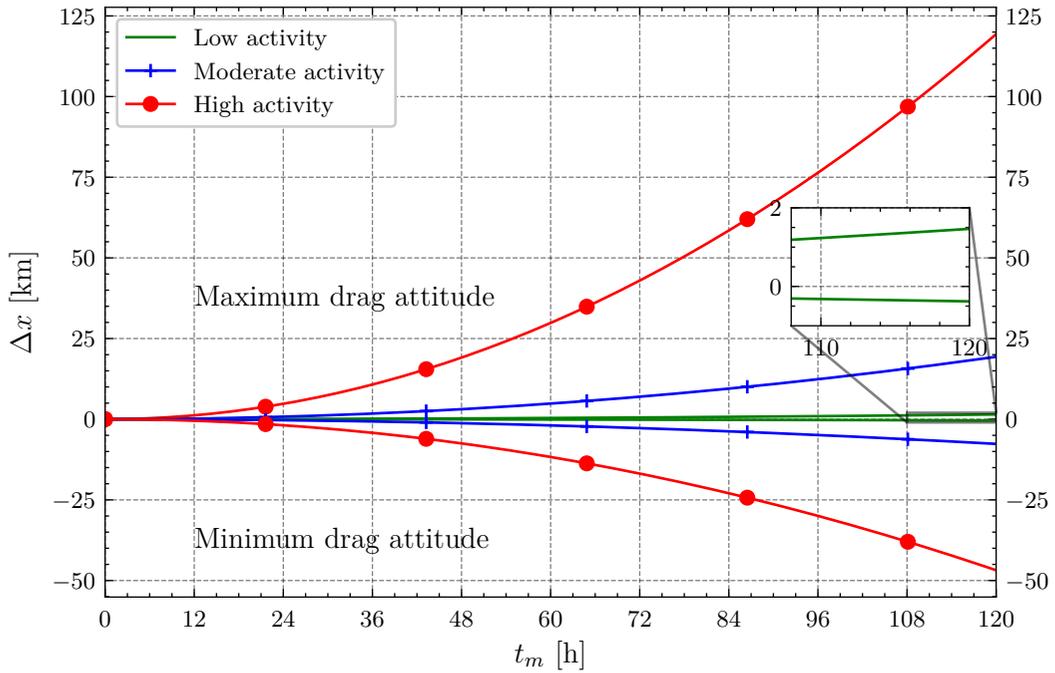}
	\caption{Achievable separation distance $ \Delta x $ of the \flp for different levels of solar and geomagnetic activity \cite{Turco.2023}.}
	\label{fig:analysis_sep_dist}
\end{figure*}

The resulting separation distances over the manoeuvring time $ t_m $ are presented in \cref{fig:analysis_sep_dist}. A flight with increased IBC leads to a positive separation distance, while a decreased IBC allows for the build-up of negative separation distances.
For low solar and geomagnetic activity, the resulting low density causes the separation distance to turn out negligibly small at only $ \qty{1.465}{\kilo\meter} $ for a flight in maximum drag attitude after $ \qty{120}{\hour} $ and even smaller for a flight with minimum drag. At moderate activity, the separation distances are noticeably higher at $ \qty{19.35}{\kilo\meter} $ and  $ \qty{-7.647}{\kilo\meter} $ after $ \qty{120}{\hour} $, respectively. For a high level of activity, the separation distances grow up to $ \qty{119.4}{\kilo\meter} $ and  $ \qty{-46.81}{\kilo\meter} $ in the same time.\par

The achievable separation distance in a given time is strongly influenced by the activity level, leading to separation distances which vary across two orders of magnitude.
Mishne and Edlerman \cite{Mishne.2017} argued that for a reasonable collision avoidance manoeuvre, the achievable separation distance must be in the range of $ \qty{900}{\meter} $ per three days for it to be greater than typical propagation uncertainties and thus useful. It can be concluded that the \flp can achieve this separation distance for moderate and high levels of solar and geomagnetic activity, both in minimum and maximum drag manoeuvres. Only for low activity, the resulting forces due to aerodynamic drag appear to be too low for creating significant separations. The results in \cref{sec:cdm_analysis} showed that, on average, the available time from reception of a warning to TCA is higher than three days and sufficient for the implementation of an aerodynamic CAM.

Solar activity follows approximately an 11-year cycle and the next maximum is expected for 2025 \cite{Hathaway.2015, NOAA.2019}. 
This allows concluding that the feasibility of aerodynamic CAMs will be given in the upcoming period until the next minimum in the solar cycle. During minimum phases, CAMs using aerodynamic drag might temporarily not be feasible for the \flp, as long as solar activity and consequently atmospheric density are low.

It is to note, that these results strongly depend on the ballistic coefficient. The used IBCs are considered lower limits. Hence, the achievable separation distances are lower limits as well. The absolute separation distance might be smaller for a minimum drag manoeuvre. Here, further analysis of the error in the determined IBC is necessary.
\subsection{Comparison with flight data}
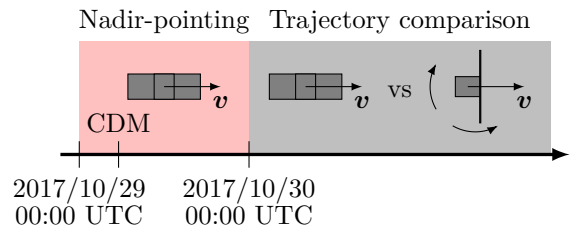
\begin{figure}[b]
	\centering
	\begin{tikzpicture}
		\fill[fill=red, text=black, opacity=0.25] (0.25,0) rectangle ++(2.25,1.5);
		\fill[fill=gray, text=black, opacity=0.5] (2.5,0) rectangle ++(4,1.5);
		
		\draw[line width=1.25, arrows={-latex}](0.0,0)--(6.75,0);
		
		\foreach \x/\xtext in {0.25/\umbruch{2017/10/29\\00:00 UTC},2.5/\umbruch{2017/10/30\\00:00 UTC}} \draw(\x,5pt)--(\x,-5pt) node[below] {\xtext};
		\foreach \x/\xtext in {0.7708/CDM} \draw(\x,-5pt)--(\x,5pt) node[above] {\xtext};
		\node[below=0.4ex, text=black] at (1.375,2.1) {Nadir-pointing};
		\node[below=0.4ex, text=black] at (4.5,2.1) {Trajectory comparison};
		
		\node[rectangle, draw, fill=gray, opacity=1, minimum width=1.922cm*0.5, minimum height=0.67cm*0.5] at (0+1.375,0+0.9) {};
		\node[rectangle, draw, fill=gray, opacity=1, minimum width=0.53cm*0.5, minimum height=0.67cm*0.5] at (0+1.375,0+0.9) {};
		\draw[arrows={-latex}]  (0+1.375,0+0.9) -- (0.75+1.375,0+0.9) node[anchor=north]{$\v{v}$};

		\node[below=0.4ex, text=black] at (4.5,1.1) {vs};
		\node[rectangle, draw, fill=gray, opacity=1, minimum width=1.922cm*0.5, minimum height=0.67cm*0.5] at (0+3.25,0+0.9) {};
		\node[rectangle, draw, fill=gray, opacity=1, minimum width=0.53cm*0.5, minimum height=0.67cm*0.5] at (0+3.25,0+0.9) {};
		\draw[arrows={-latex}]  (0+3.25,0+0.9) -- (0.75+3.25,0+0.9) node[anchor=north]{$\v{v}$};
		
		\draw[fill=gray, opacity=1, minimum width=0.66cm*0.5, minimum height=0.53cm*0.5]{} (0-0.33*0.5+5.4,-0.265*0.5+0.9) rectangle ++(0.66*0.5,0.53*0.5);
		\draw[fill=gray, opacity=1, minimum width=0.03cm*0.5, minimum height=1.922cm*0.5]{} (0.33*0.5-0.015*0.5+5.4,-0.961*0.5+0.9) rectangle ++(0.03*0.5,1.922*0.5);
		\draw[arrows={-latex}]  (0+5.4,0+0.9) -- (0.75+5.4,0+0.9) node[anchor=north]{$\v{v}$};
		\draw [-latex, black, domain=270-30:270+30] plot ({5.5+0.6*cos(\x)}, {0.9+0.6*sin(\x)});
		\draw [-latex, black, domain=180+30:180-30] plot ({5.5+0.6*cos(\x)}, {0.9+0.6*sin(\x)});
		
	\end{tikzpicture}
	\caption{Timeline of the flight test.}
	\label{fig:timeline}
\end{figure}

On October 29, 2017, the \flp performed a 24-hour Nadir-pointing before going back to normal operation, \ie not holding a specified attitude for a longer duration. This will be used as a test case for the analytic estimation of the separation distance.
The change in attitude at the end of the Nadir-pointing will lead to a separation distance $ \Delta x $ compared to a reference trajectory of the satellite flying further in Nadir-pointing. This separation distance will in the following be compared to the analytic estimation. The timeline of the flight test is presented in \cref{fig:timeline}.\par
During the Nadir-pointing, a two-line element set (TLE) has been created by JSpOC with epoch October 29, 2017, 09:58:17 UTC, containing information about the orbit for further propagation. The propagated positions of this TLE will be used as a reference trajectory to compare it with GPS data over a course of seven days. The IBC of the satellite in Nadir-pointing attitude has been determined in \cref{sec:aero_analysis}. The IBC of the flight phase after the Nadir-pointing is taken from a CDM for a close encounter on November 13, 2017, 
and set to be $ \beta^* = 0.02373 $. Based on these IBCs, the presented tool is used to estimate the achievable separation distance. 
At the same time, GPS positions from the \flp are compared to the sgp4-propagated reference trajectory and their in-track separation is determined.\par

\begin{figure*}[p!]	
	\centering
	\input{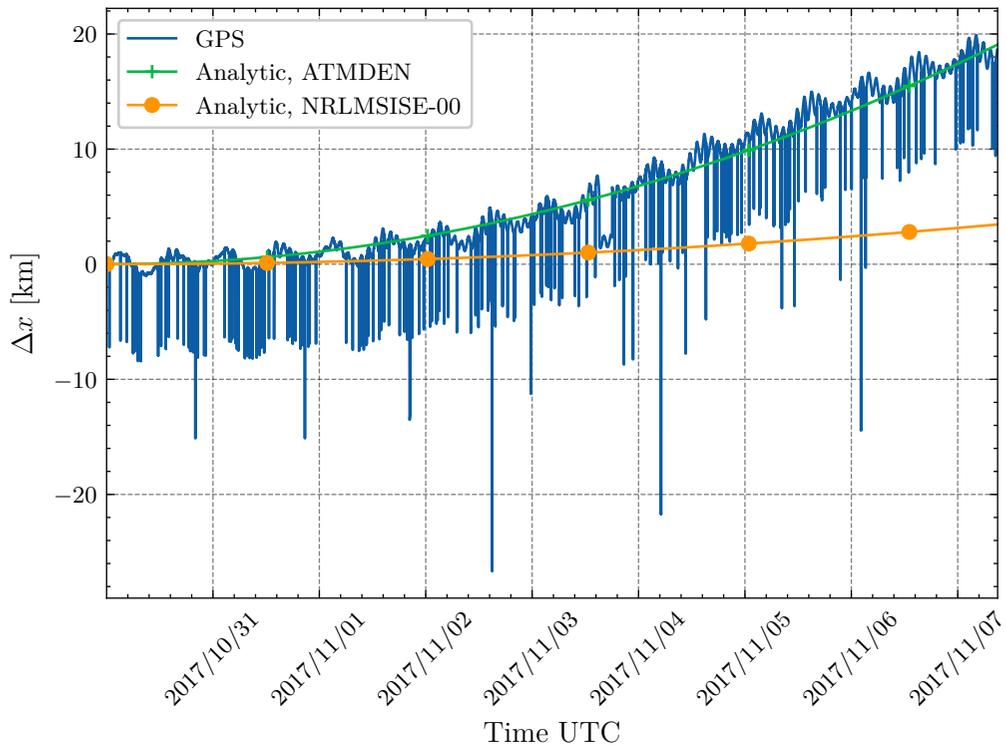}
	\caption{Comparison of the separation distance $ \Delta x $ relative to the reference trajectory obtained through GPS data and sgp4-propagation of a TLE and through the analytic equation for a flight test case in October 2017.}
	\label{fig:flight_test_Nadir}
\end{figure*}

\cref{fig:flight_test_Nadir} shows the resulting separation distances building up after the end of the Nadir-pointing. While the difference between GPS positions and the propagated orbit of the TLE is noisy, a separation distance in positive direction can be seen, growing quadratically to almost $ \SI{20}{\kilo\meter} $ in the observed seven days. The analytically estimated separation making use of the NRLMSISE-00 atmosphere model is significantly smaller than the separation between GPS positions and sgp4-propagation, measuring only $ \SI{3.453}{\kilo\meter} $. It was, however, found that the \emph{Atmospheric Density Estimate for Drag Calculation} (ATMDEN) service forecasted a higher density for the observed period. The service employs a drag temperature model for estimating atmospheric densities and issues density estimates depending on height, as well as latitude and longitude. Using the obtained density values by ATMDEN for October 31, 2017, the separation distance estimated by the tool fits well to the observed separation between GPS positions and sgp4-propagation.\par
While the general effect of a variation of the ballistic coefficient is visible in the flight data, a further flight test will be necessary to quantify the accuracy of the separation distance estimation. It is advised that such a flight test may be performed over the course of several days: The satellite should hold a defined attitude, \eg minimum drag, during an orbit determination process before changing its attitude to the manoeuvring attitude, \eg maximum drag, for a duration of several days, resembling a realistic CAM. The effect of the manoeuvre is to be assessed by analysing the GPS data of the satellite.

\section{Summary}\label{sec13}
This work proved the feasibility of collision avoidance manoeuvres using aerodynamic drag for the \flp satellite for moderate and high solar and geomagnetic activity. The satellite is subject to recurring close encounter warnings but has no thrusters and, therefore, cannot perform impulsive evasive manoeuvres. \par
The necessary steps included an aerodynamic analysis of the satellite, during which the inverse ballistic coefficient of a simplified CAD model could be estimated for attitudes maximizing and minimizing the aerodynamic drag force on the satellite, respectively. The coefficients showed a dependency on solar and geomagnetic activity. An analysis of previous collision warnings showed that on average collision warnings are issued early enough to implement evasive manoeuvres. During moderate and high solar and geomagnetic activity, the \flp can achieve sufficient separation distances, as was shown using an analysis tool based on an analytic equation from literature.\par
Lastly, the results of the analytically estimated separation distance were compared to flight data. While the general effect on the trajectory, \ie an in-plane separation, becomes visible, a dedicated flight test in the future will be necessary to draw stronger conclusions.\par

\section{Outlook}\label{sec14}
Regarding the \flp, the aerodynamic analysis is of great importance to future work. Analyses with Direct Simulation Monte Carlo Methods may help to better assess the quality of the determined coefficients. Additionally, a further flight test with the \flp is expected to deliver useful insights.\par
Another focus of future work will be the development of a general CAM analysis tool. While the feasibility of aerodynamic CAMs was shown for the \flp, the tool may also be used for other satellites. Satellites in similar and especially in lower orbits which can vary their ballistic coefficient can benefit from aerodynamic collision avoidance manoeuvres as well to decrease the risk of potential collisions.

\backmatter

%
%

%

\section*{Declarations}
%
\setlist[description]{font=\rmfamily\upshape\bfseries}
\begin{description}
\item [Conflict of interest] On behalf of all authors, the corresponding author states that there is no conflict of interest. 
\end{description}
%

%
%
%
%

\begin{appendices}


\section{Space weather parameters}
\begin{table}[h]
	\centering
	\caption{Levels of solar and geomagnetic activity as defined in ISO1422 \cite{ISO1422}. Each value is given in the respective index unit and is to be applied for the average index value as well.}
	\label{tab:activity_levels}
	\begin{tabular}{@{}lcccccc@{}}
		\toprule
		Activity level &$F_{10.7}$ &$S_{10}$ &$M_{10}$ &$Y_{10}$ &$A_p$ &$Dst$\\
		\midrule
		Low &65 &60 &60 &60 &0 &-15\\
		Moderate &140 &125 &125 &125 &15 &-15\\
		High &250 &220 &220 &220 &45 &-100\\
		\bottomrule
	\end{tabular}
\end{table}

\end{appendices}


\setlength{\bibsep}{0pt plus 0.3ex}
\bibliography{sn-bibliography}


\end{document}